\documentclass[aps,showpacs,preprintnumbers,amsmath, amssymb]{revtex4}

\usepackage{float}
\usepackage{graphics,epsfig}
\usepackage{graphicx}
\usepackage{dcolumn}
\usepackage{bm}
\usepackage{amsmath}

\begin{document}
\baselineskip=0.8 cm

\title{{\bf On instabilities of stationary scalar field configurations supported by  reflecting compact stars}}
\author{Yan Peng$^{1}$\footnote{yanpengphy@163.com}}
\affiliation{\\$^{1}$ School of Mathematical Sciences, Qufu Normal University, Qufu, Shandong 273165, China}

\vspace*{0.2cm}
\begin{abstract}
\baselineskip=0.6 cm
\begin{center}
{\bf Abstract}
\end{center}

We study instabilities of the system composed of
stationary scalar fields and asymptotically
flat horizonless reflecting compact stars.
In the probe limit, we obtain bounds on the
scalar field frequency. Below this bound, stationary hairy
stars are expected to suffer from nonlinear instabilities
under massless field perturbations.
In other words, we prove that stationary scalar hairy stars are unstable
for scalar fields with small frequency.

\end{abstract}

\maketitle
\newpage
\vspace*{0.2cm}

\section{Introduction}

There is accumulating evidence that fundamental scalar fields may exist in nature \cite{Franz}.
Theoretically, the scalar field can be either static or stationary.
The famous black hole no scalar hair theorems
state an interesting property that static scalar hair usually cannot exist in the asymptotically flat
black hole backgrounds, see references \cite{Bekenstein}-\cite{Bar} and reviews \cite{Bekenstein-1,CAR}.
In contrast, it has recently been shown that rotating black holes allow the existence of stationary massive
scalar field hairs \cite{Hod-1}-\cite{st11}.
Moreover, with scalar fields confined in a box, static hairy black hole solutions were constructed in \cite{Dolan,Basu,Oscar}
and their dynamical formation was studied in \cite{Sanchis}.
We should mention that the instability properties of hairy black holes were investigated \cite{sta1,sta2}.

Interestingly, no static scalar hair behavior also appears in horizonless neutral
compact object spacetimes. Hod firstly proved no static scalar hair theorems for asymptotically
flat horizonless neutral compact stars with reflective surface boundary conditions \cite{Hod-6}.
In fact, being filled with matter, it is natural to assume that the
compact star surface would have reflective properties \cite{bg}.
When considering nonminimal couplings between scalar fields and curvature,
static scalar hairs also cannot exist outside asymptotically
flat horizonless neutral reflecting compact stars \cite{Hod-7}.
In the asymptotically dS background, no static scalar hair theorems still hold
for the horizonless neutral reflecting compact star \cite{Bhattacharjee}.
So the no static scalar hair behavior is a very general property in
the spacetime of horizonless neutral reflecting compact stars.

On the other side, null circular geodesics may exist
in the compact object spacetime \cite{P1,P2,P3,P4,P5}.
For horizonless compact objects,
if null circular geodesics exist, in general, there will be
pairs of the null circular geodesics and
the innermost null circular geodesic is stable \cite{IN1,IN2}.
So horizonless compact objects with null circular geodesics
is expected to be unstable since massless fields
can pile up on the innermost stable null geodesic \cite{IS1,IS2}.
One known way to evade the no hair theorem for horizonless reflecting compact star
is to consider a charged background \cite{Hod-8}-\cite{YP-5}.
The null circular geodesic was used to study instabilities
of charged horizonless static scalar hairy compact stars \cite{ISPY,ISP}.
Recently, Hod provided another interesting way to evade the
no hair theorem for horizonless reflecting compact stars, which is considering stationary
scalar field hairs \cite{SH}. So it is interesting to disclose
the (in)stability of such horizonless stationary scalar hairy reflecting compact stars
through properties of null circular geodesics.

In the following, we study the (in)stability of horizonless stationary scalar hairy
reflecting compact stars in the asymptotically flat gravity.
We analytically obtain bounds for the scalar field frequency.
Below this bound, stationary hairy stars are unstable.
We give conclusions at the last section.

\section{Bounds for the frequency of stationary scalar fields}

We study a gravity model of stationary scalar fields
linearly coupled to horizonless reflecting compact stars.
And the matter field Lagrange density is
\begin{eqnarray}\label{lagrange-1}
\mathcal{L}=-|\nabla_{\mu} \Psi|^{2}-m^{2}\Psi^{2},
\end{eqnarray}
where $\Psi$ is the scalar field with mass m.

The line element of the spherically symmetric compact star reads
\cite{P1}
\begin{eqnarray}\label{AdSBH}
ds^{2}&=&-fe^{-\chi}dt^{2}+\frac{dr^{2}}{f}+r^{2}(d\theta^{2}+sin^{2}\theta d\phi^{2}).
\end{eqnarray}
Here $f$ and $\chi$ are metric functions depending on the radial coordinate r.
We define the star radius as $r_{s}$.
Since the spacetime is horizonless,
there is $f(r)>0$ for all $r\geqslant r_{s}$.
$\theta$ and $\phi$ are angular coordinates.

The equations of metrics and scalar fields are \cite{ISP,ISPY,SH,dyp,metric1,metric2,metric3}
\begin{eqnarray}\label{BHg}
f'=-8\pi r \rho+(1-f)/r,
\end{eqnarray}
\begin{eqnarray}\label{BHg}
\chi'=-8\pi r (\rho+p)/f,
\end{eqnarray}
\begin{eqnarray}\label{BHg}
(\nabla^{\nu}\nabla_{\nu}-m^2)\Psi=0
\end{eqnarray}
with $\rho=-T^{t}_{t}$, $p=T^{r}_{r}$ as the
matter field energy density and the radial pressure respectively.

In this work, we neglect scalar fields' backreaction on the background.
So there is Schwarzschild type solution $\chi(r)=0$ and $f(r)=1-\frac{2M}{r}$
with $M$ as the star mass.
We take stationary scalar fields in the form
\begin{eqnarray}\label{BHg}
\Psi(t,r)=e^{-i \omega t}\psi(r),
\end{eqnarray}
where $\omega$ is the frequency.

And the scalar field equation is
\begin{eqnarray}\label{BHg}
\psi''+(\frac{2}{r}+\frac{f'}{f})\psi'+(\frac{\omega^2}{f^2}-\frac{m^2}{f})\psi=0
\end{eqnarray}
with $f=1-\frac{2M}{r}$ \cite{blc1,blc2,blc3,blc4,blc5}.

At the star surface $r_{s}$, we impose the scalar reflecting condition.
At the infinity, the general asymptotic behavior
is $\psi\sim A\cdot \frac{1}{r}e^{-\sqrt{m^2-\omega^2}r}
+B\cdot \frac{1}{r}e^{\sqrt{m^2-\omega^2}r}$,
where A and B are integral constants.
Boundness of the scalar field at infinity requires $B=0$ \cite{BC}.
So boundary conditions are
\begin{eqnarray}\label{InfBH}
&&\psi(r_{s})=0,~~~~~~~~~~~~~\psi(\infty)=0.
\end{eqnarray}

It was shown that horizonless compact objects with null circular geodesics
usually have pairs of null circular geodesics
and the innermost null circular geodesic is stable \cite{IN1}.
As massless fields tend to pile up on the stable null circular geodesic,
horizonless compact stars with null circular geodesics are
unstable to massless field perturbations \cite{IS1,IS2}.
So we can study the instability of horizonless stationary hairy configurations
by examining whether there is null circular geodesic in the spacetime.
In the probe limit, there is exterior null circular geodesic
when the would-be null circular geodesic radius is above
compact star surface. In the following analysis, we will obtain the
instability condition by imposing exterior null circular geodesic radii
above upper bounds of hairy star radii.

We introduce a new function $\tilde{\psi}=\sqrt{r}\psi$.
According to the scalar field equation (7), the equation of
the function $\tilde{\psi}$ can be expressed as
\begin{eqnarray}\label{BHg}
r^2\tilde{\psi}''+(r+\frac{r^2f'}{f})\tilde{\psi}'+(-\frac{1}{4}-\frac{rf'}{2f}+\frac{\omega^2r^2}{f^2}-\frac{m^2r^2}{f})\tilde{\psi}=0
\end{eqnarray}
with $f=1-\frac{2M}{r}$.

Boundary conditions of the function $\tilde{\psi}$ are
\begin{eqnarray}\label{InfBH}
&&\tilde{\psi}(r_{s})=0,~~~~~~~~~\tilde{\psi}(\infty)=0.
\end{eqnarray}

According to (10), at least one extremum
point $r=r_{peak}$ of the function $\tilde{\psi}$
exists between the surface $r_{s}$ and the infinity. At this extremum point, there are the following relations
\begin{eqnarray}\label{InfBH}
\{ \tilde{\psi}'=0~~~~and~~~~\tilde{\psi} \tilde{\psi}''\leqslant0\}~~~~for~~~~r=r_{peak}.
\end{eqnarray}

Relations (9) and (11) yield the inequality
\begin{eqnarray}\label{BHg}
-\frac{1}{4}-\frac{rf'}{2f}+\frac{\omega^2r^2}{f^2}-\frac{m^2r^2}{f}\geqslant0~~~for~~~r=r_{peak}.
\end{eqnarray}

It can be transformed into
\begin{eqnarray}\label{BHg}
m^2r^2f(r)\leqslant \omega^2r^2-\frac{rff'}{2}-\frac{1}{4}f^2~~~for~~~r=r_{peak}.
\end{eqnarray}

The regular condition of the spacetime requires
$r_{s}\geqslant 2M$, otherwise there is a horizon
at $r=2M$ above the star surface.
With $r_{s}\geqslant 2M$, there are relations
\begin{eqnarray}\label{BHg}
r\geqslant r_{s}\geqslant 2M,
\end{eqnarray}
\begin{eqnarray}\label{BHg}
f=1-\frac{2M}{r}=\frac{1}{r}(r-2M)\geqslant0,
\end{eqnarray}
\begin{eqnarray}\label{BHg}
rf'=r(\frac{2M}{r^2})=\frac{2M}{r} \geqslant 0.
\end{eqnarray}

According to (13-16), the following inequality holds
\begin{eqnarray}\label{BHg}
m^2r^2f(r)\leqslant \omega^2r^2~~~for~~~r=r_{peak}.
\end{eqnarray}

The relation (17) yields the inequality
\begin{eqnarray}\label{BHg}
m^2r^2(1-\frac{2M}{r})\leqslant \omega^2r^2~~~for~~~r=r_{peak}.
\end{eqnarray}

We can transform (18) into
\begin{eqnarray}\label{BHg}
m^2-\omega^2\leqslant \frac{2m^2M}{r_{peak}}.
\end{eqnarray}

From (19), we obtain bounds on the extremum
point $r=r_{peak}$ in the form
\begin{eqnarray}\label{BHg}
r_{peak}\leqslant \frac{2m^2M}{m^2-\omega^2}.
\end{eqnarray}

It is also the bound on hairy star radii as
\begin{eqnarray}\label{BHg}
r_{s}\leqslant r_{peak}\leqslant \frac{2m^2M}{m^2-\omega^2}.
\end{eqnarray}

Following approaches in \cite{td1,td2},
we derive the characteristic relation for
null circular geodesics.
The Lagrangian describing geodesics is
\begin{eqnarray}\label{BHg}
2\mathcal{L}=-e^{-\chi}f\dot{t}^2+\frac{1}{f}\dot{r}^2+r^2\dot{\phi}^2,
\end{eqnarray}
where a dot represents a derivative with respect to the affine parameter along the geodesic.

The Lagrangian is independent of t and $\phi$.
This implies that the existence of two constants of motion labeled as E and L.
According to (22), the generalized momenta can be expressed as
\begin{eqnarray}\label{BHg}
p_{t}=-e^{-\chi}f\dot{t}=-E=const,
\end{eqnarray}
\begin{eqnarray}\label{BHg}
p_{\phi}=r^2\dot{\phi}=L=const,
\end{eqnarray}
\begin{eqnarray}\label{BHg}
p_{r}=\frac{1}{f}\dot{r}~.
\end{eqnarray}
The Hamiltonian of the system is
$\mathcal{H}=p_{t}\dot{t}+p_{r}\dot{r}+p_{\phi}\dot{\phi}-\mathcal{L}$,
which implies
\begin{eqnarray}\label{BHg}
2\mathcal{H}=-E\dot{t}+L\dot{\phi}+\frac{1}{f}\dot{r}^2=\delta=const.
\end{eqnarray}
For timelike geodesics, there is $\delta=1$ and in this paper with
null geodesics, we take $\delta=0$.

From (26), we obtain the expression
\begin{eqnarray}\label{BHg}
\dot{r}^2=f[E\dot{t}-L\dot{\phi}].
\end{eqnarray}

With relations (23) and (24), we get expressions for $\dot{t}$ and $\dot{\phi}$ in the form
\begin{eqnarray}\label{BHg}
\dot{t}=\frac{e^{\chi}E}{f},~~~~~~~~~~\dot{\phi}=\frac{L}{r^2}.
\end{eqnarray}

Putting (28) into (27), we arrive at the relation
\begin{eqnarray}\label{BHg}
\dot{r}^2=f[\frac{e^{\chi}E^2}{f}-\frac{L^2}{r^2}]=e^{\chi}E^2-\frac{L^2f}{r^2}.
\end{eqnarray}

At the null circular geodesic, there are relations $\dot{r}^2=0$ and $(\dot{r}^2)'=0$ \cite{td2}.
The equation $\dot{r}^2=0$ yields
\begin{eqnarray}\label{BHg}
E^2=\frac{L^2f}{r^2e^{\chi}}.
\end{eqnarray}

The requirement $(\dot{r}^2)'=0$ yields
\begin{eqnarray}\label{BHg}
(\dot{r}^2)'=\chi'e^{\chi}E^2+\frac{2L^2f}{r^3}-\frac{L^2f'}{r^2}=0.
\end{eqnarray}

According to (30) and (31), the characteristic equation of null circular geodesics is
\begin{eqnarray}\label{BHg}
2f(r_{\gamma})+r_{\gamma}[\chi'(r_{\gamma})f(r_{\gamma})-f'(r_{\gamma})]=0.
\end{eqnarray}

Without backreaction of scalar fields, the exterior spacetime of the compact star
can be described by $\chi=0$ and $f=1-\frac{2M}{r}$. The exterior null circular geodesic
radius is
\begin{eqnarray}\label{BHg}
r_{\gamma}=3M.
\end{eqnarray}

In order to obtain the instability condition,
we impose that the radius of the would-be outer null circular geodesics
is above the hairy star radius bound (21),
which leads to the existence of the exterior null circular geodesic $r_{\gamma}=3M$ \cite{ISP,ISPY}.
So the horizonless hairy star is unstable on
condition that
\begin{eqnarray}\label{BHg}
r_{\gamma}=3M\geqslant \frac{2m^2M}{m^2-\omega^2}.
\end{eqnarray}

From (34), we obtain bounds for the frequency of stationary scalar fields as
\begin{eqnarray}\label{BHg}
\frac{\omega^2}{m^2}\leqslant \frac{1}{3}.
\end{eqnarray}
It is known that the static ($\omega=0$) scalar field usually cannot exist outside
horizonless neutral reflecting compact stars. In contrast, stationary ($\omega\neq0$)
scalar fields can condense outside the horizonless neutral reflecting compact star.
In this work, we further show that horizonless neutral stationary hairy reflecting
compact stars are unstable for frequency below the bound (35).

\section{Conclusions}

We investigated the model of stationary scalar fields linearly coupled to
asymptotically flat horizonless neutral compact stars
with reflecting boundary conditions.
We studied instabilities of scalar hairy reflecting stars
through properties of null circular geodesics.
We analytically obtained bounds on the frequency of stationary scalar field
as $\frac{\omega^2}{m^2}\leqslant \frac{1}{3}$,
where $\omega$ and m are the frequency and mass of the scalar field respectively.
Below this bound, stationary scalar hairy configurations
supported by horizonless reflecting compact stars are expected to be dynamically unstable
under perturbations of massless fields.
That is to say the horizonless stationary scalar hairy compact reflecting star
is unstable for scalar fields with small frequency.

\begin{acknowledgments}

This work was supported by the Shandong Provincial Natural Science Foundation of China under Grant
No. ZR2018QA008. This work was also supported by a grant from Qufu Normal University of China under Grant
No. xkjjc201906.

\end{acknowledgments}

\end{document}